\documentclass[prl,twocolumn,showpacs,preprintnumbers,superscriptaddress,amsmath,amssymb]{revtex4}

\usepackage{epsf,color,graphicx}
\usepackage[dvips]{epsfig}

\begin{document}
\title{Stimulated and spontaneous four-wave mixing in silicon-on-insulator coupled photonic wire nano-cavities}

\author{Stefano Azzini}
\author{Davide Grassani}
\author{Matteo Galli}
\author{Dario Gerace}
\author{Maddalena Patrini}
\author{Marco Liscidini}
\affiliation{Dipartimento di Fisica, Universit\`{a} degli Studi di Pavia, via Bassi 6, 27100 Pavia, Italy}
\author{Philippe Velha}
\affiliation{School of Engineering, University of Glasgow, Glasgow G12 8LT, UK}
\author{Daniele Bajoni}\email[]{daniele.bajoni@unipv.it}
\affiliation{Dipartimento di Ingegneria Industriale e dell'Informazione, Universit\`{a} degli Studi di Pavia, via Ferrata 1, Pavia, Italy}

\date{\today}

\begin{abstract}
We report on four-wave mixing in coupled photonic crystal nano-cavities on a silicon-on-insulator platform. Three photonic wire cavities are side-coupled to obtain three modes equally separated in energy. The structure is designed to be self-filtering, and we show that the pump is rejected by almost two orders of magnitudes. We study both the stimulated and the spontaneous four-wave mixing processes: owing to the small modal volume, we find that signal and idler photons are generated with a hundred-fold increase in efficiency as compared to silicon micro-ring resonators.
\end{abstract}


\maketitle


The adoption of photonic devices integrated on semiconductor chips is increasing over the years. Circuits containing light sources, waveguides, and detectors have been successfully demonstrated and are used to route information between components in high-end computer systems as well as in data centers. Silicon is by far the mostly used material in a wide range of technological processes. Due to its strong refractive index contrast with air and silica, silicon-on-insulator (SOI) wafers can be processed in devices for efficient confinement and propagation of light, such as photonic wire (PhW) waveguides. The goal of obtaining active functionalities in such devices has started a thorough investigation on silicon optical nonlinearities \cite{Wynne69}. The inversion symmetry of the crystalline unit cell prohibits second-order nonlinearities in bulk silicon, but third order nonlinear effects have been reported, including Raman scattering \cite{Ralston70,Claps02}, Kerr nonlinearity \cite{Dinu03,Tsang02} and two-photon absorption \cite{Reintjes73}.

\begin{figure}[b!]
\includegraphics[width= \columnwidth]{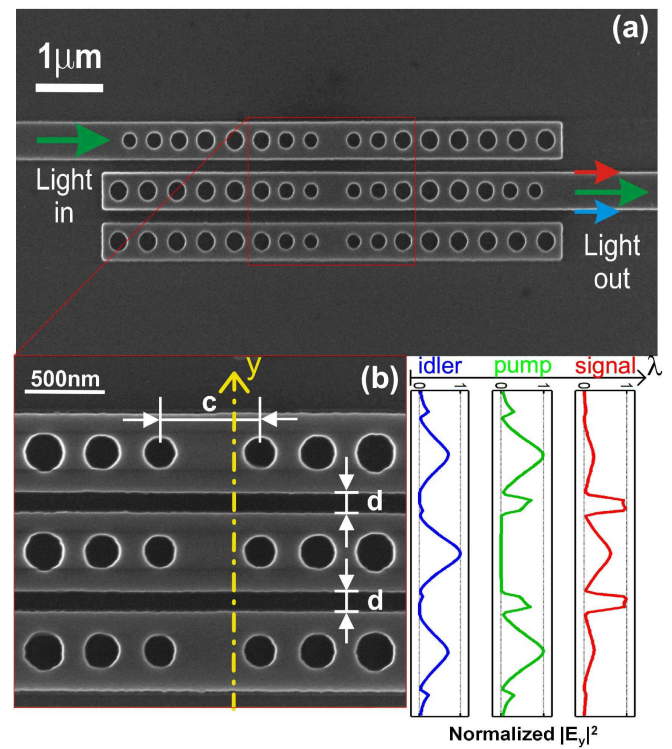}
\caption{(color online) (a) SEM image of the SOI-integrated photonic crystal molecule. (b) Left: SEM close-up view of the triple cavity region, showing the cavity length c$=$620nm and the coupling distance d$=$100nm. Right: FDTD-calculated electric field intensity cross-section of the three eigenmodes of the coupled structure along the dash-dotted yellow line (y-direction).}\label{Fig1}
\end{figure}

A common way to enhance optical nonlinearities is to use optical confinement to increase the overlap between the electromagnetic field and the nonlinear material. In these regards, photonic crystals (PhCs) \cite{Joannopoulos}, and in particular PhC nano-cavities \cite{Noda}, allow for the confinement of light to volumes comparable to the cube of their wavelength in the material and can be fabricated using several top-down approaches \cite{CMOS1,CMOS2} compatible with CMOS technology \cite{CMOS3,CMOS4}. PhC cavities have been employed to enhance basic cavity quantum electrodynamics effects \cite{Badolato,Englund}, to demonstrate ultra-low threshold lasing \cite{Hennessy,L3pollaser} and to enhance optical nonlinear effects in silicon \cite{Notomi1, Notomi2, Belotti,GalliTHG}.

In this work we use a PhC molecule (PCM) to achieve four-wave mixing (FWM) in a silicon integrated device. In order to achieve the energy conservation necessary for FWM, three resonances evenly separated in energy are necessary \cite{Turner08}. One way to achieve this condition is to exploit the natural splitting in the resonances ot three coupled identical cavities: this was first demonstrated for one dimensional cavities in strong coupling with quantum well excitons \cite{Tignon}, but the same geometry can work with three dimensional cavities in any material with a $\chi^{(3)}$ optical nonlinearity.

Our sample consists of three nominally-identical side-coupled PhC/PhW nano-cavities on SOI platform \cite{wire1,wire2}, as shown in the scanning electron micrograph of the integrated device reported in Fig. \ref{Fig1}(a). The starting point for our triple cavity design is our previously reported PhW nano-cavity structure \cite{Belotti}. We use here 220nm-thick and 500nm-wide silicon ridge waveguides, supporting a single transverse electric mode. The PhC is patterned as a 1D linear array of air-holes of radius \emph{r} with lattice constant a$=$420 nm and $r/a=0.27$, introducing, in the guided mode band structure, a $\sim$350 nm-wide transmission stop-band centred at 1550 nm. At the middle of the PhC structure, a cavity-defect is introduced, whose length \emph{c} is 620nm from centre-to-centre of the two inner holes. The cavity mirrors are tapered nearby the cavity-defect (see Fig. \ref{Fig1}(b)), namely the holes size and spacing are modified to allow for adiabatic modal conversion \cite{Lalanne}, and thus granting for strong light localization. The triple cavity structure of our PCM is created by the strong evanescent coupling of three such cavities placed side-by-side at a coupling distance \emph{d} of 100 nm. The whole structure is very compact: the footprint of the device is about 10$\mu$m$^{2}$. The samples were fabricated using commercial SOI wafers from SOITEC by means of state-of-the-art nano-fabrication techniques, involving electron beam lithography (EBL) and dry-etching. A negative-tone e-beam resist, a 1:1 dilution of hydrogen silsesquioxane (HSQ) in metil isobutyl ketone (MIBK), was spun onto the SOI sample at 3000 rpm for 60 seconds and baked at 90$°$C for 2 minutes, to give an approximatively 200 nm-thick layer. The photonic patterns were then directly written in the HSQ layer by using a Vistec VB6 EBL tool operated at 100keV, and, after development, they were finally transferred into the silicon guiding layer by means of an inductively coupled plasma reactive ion etching process employing a combined SF$_{6}$/C$_{4}$F$_{8}$ gas chemistry, in order to obtain waveguide structures with smooth side-walls. Spot-size converters comprising an inverse taper of the silicon waveguide covered by a micron-sized polymer waveguide were also fabricated to get an efficient input and output coupling to the photonic nano-wires.

\begin{figure}[t!]
\includegraphics[width= \columnwidth]{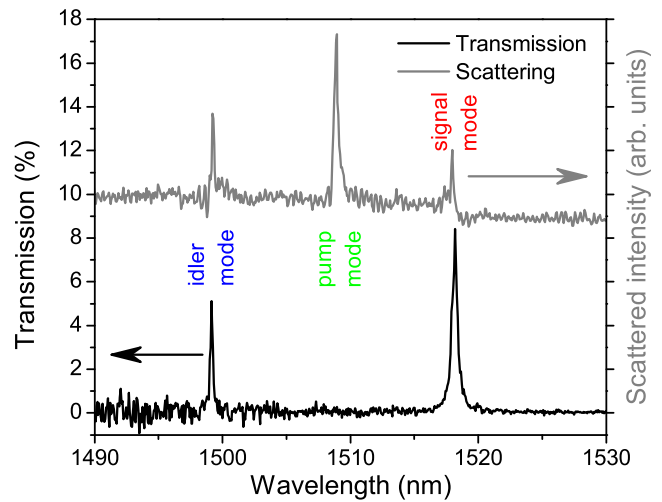}
\caption{(color online) Transmission spectrum (black line, left axis) and scattering spectrum (grey line, right axis) from the sample. The scattering spectrum has been vertically shifted for clarity.}\label{Fig2}
\end{figure}

An end-fire coupling experimental setup is used for both linear characterization and four-wave mixing experiments. The PCM was characterized using a broadband incoherent light source (SLED, 200nm bandwidth centred at 1550nm) and a spectrometer coupled to a liquid nitrogen cooled InGaAs CCD array for detection: a transmission spectrum of the device is shown as a black curve in Fig. \ref{Fig2}. As it was expected from numerical simulations (see Fig. \ref{Fig1}(b)), collecting the light from the central cavity effectively allows to reject the central (pump) mode, while the blue (idler) and red (signal) modes are transmitted. Therefore, the triple cavity structure is self-filtering in transmission, and we measured a pump intensity rejection of almost two orders of magnitude. This is beneficial in the case of transmission FWM experiments, in particular in the spontaneous FWM, as the pump intensity is typically several orders of magnitude stronger than the intensities of the generated signal and idler beams. 

To prove efficient coupling of the pump beam with the photonic molecule, we detected the light scattered from the top of the sample by collecting it with a high numerical aperture microscope objective. As shown by the grey curve in Fig. \ref{Fig2}, the three resonances due to the light scattered by the cavities are clearly visible on a small background given by the scattering contribution coming from the input waveguide \cite{Fano}. The idler, pump, and signal modes have, respectively, wavelengths of 1499.1 nm, 1508.9 nm, and 1518.2 nm with quality factors of $\sim$6000, $\sim$4000, and $\sim$3000. The average energy between the blue and the red modes falls at 1508.7 nm, well within the FWHM of the central mode, so that energy conservation for FWM is satisfied. 

We notice that, in the case of our PCM, coupled-mode theory can offer only a qualitative description of the spectral positions of the three resonant modes \cite{Peyrademolecule}. Indeed, as the PhC nano-cavities are very close to each other (d$=$100nm), the assumption of weakly coupled resonators  is no longer valid, and the theory predicts an asymmetric mode splitting in the case of nominally identical coupled-cavities \cite{CIFS,Whittakermolecule}. However, fabrication imperfections can compensate this effect and bring about the three modes to be evenly separated in energy, as it is for the device reported in this paper. This is consistent with the fact that a dielectric cavity resonance frequency is expected to be sensitive to small modifications in dimensions or refractive index \cite{CIFS}. 

For the nonlinear measurements, two continuous wave TSL-510 Santec lasers are used, one set at the pump wavelength, and the other one set at the signal wavelength in the case of the stimulated experiment. The lasers are spectrally cleaned using bandpass filters, and other bandpass filters are also used to separate the generated signal and idler beams from the pump; more details on the experimental setup can be found in Refs. \cite{OLrings,OErings}.
\begin{figure}[t!]
\includegraphics[width= \columnwidth]{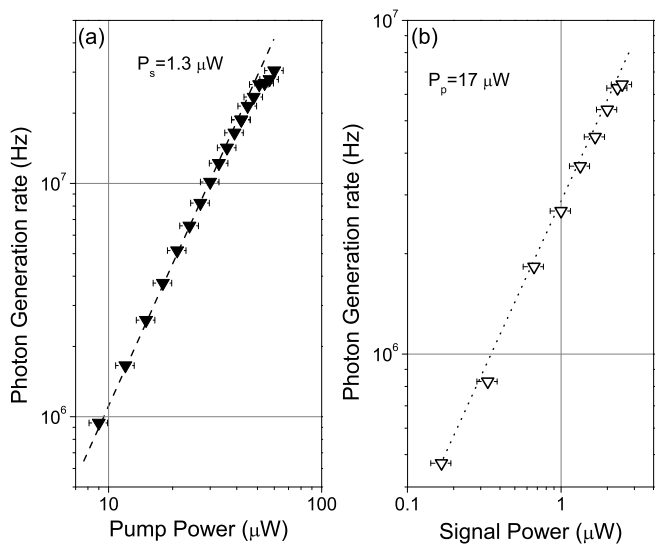}
\caption{(color online) Scaling of the estimated number of generated  photons for stimulated FWM with varying (a) the pump power (the dashed line is a guide to the eye proportional to the square of the pump power), and (b) the input signal power with 17$\mu$W injected at the pump resonance (the dashed line is a guide to the eye proportional to the signal power). }\label{Fig3}
\end{figure}
In Fig. \ref{Fig3} we report the integrated intensity of the idler beam generated in the stimulated FWM as a function of the pump power at a fixed signal input power P$_{s}$=1.3$\mu$W (Fig. \ref{Fig3}(a)), and as a function of the signal power at a fixed pump power P$_{p}$=17$\mu$W (Fig. \ref{Fig3}(b)). The estimated photon generation rate exhibits a clear quadratic scaling in the former case and a linear scaling in the latter, a proof that the process is indeed due to FWM. 

\begin{figure}[t!]
\includegraphics[width= \columnwidth]{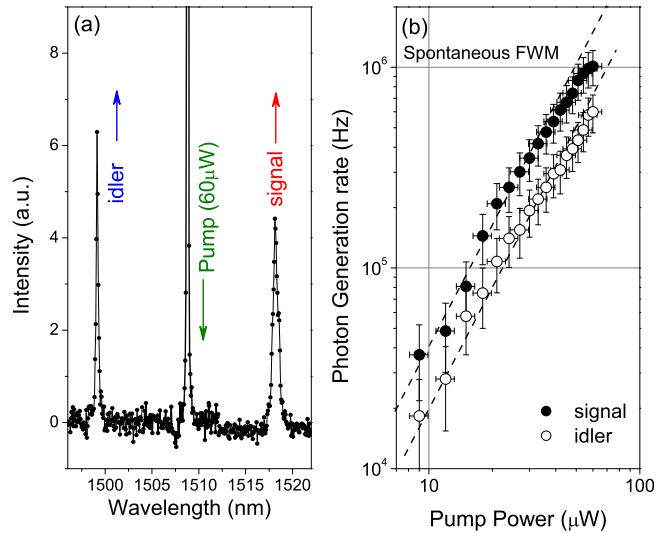}
\caption{(a) An example of a spontaneous FWM spectrum from the photonic molecule. (b) Integrated intensities of the generated beams as a function of the pump power. The lines are guides to the eye proportional to the square of the pump power.}\label{Fig4}
\end{figure}

Finally, in Fig. \ref{Fig4} we show experimental results for the spontaneous FWM experiments in which only the pump laser is coupled to the sample, and the signal and idler beams are ''stimulated" by vacuum power fluctuations. Fig. \ref{Fig4}(a) shows a typical spontaneous FWM spectrum obtained for an estimated pump power of 60 $\mu$W. In this experiments external filtering stages are used to suppress the pump, which remains too strong even considering the self-filtering properties of the cavity sample \cite{OErings}. The integrated intensities for signal and idler beams are shown in Fig. \ref{Fig4}(b).  As expected, the intensities of the spontaneously generated beams increase quadratically with the pump power. Unlike previous works reporting on the generation of photon pairs in waveguides or side coupled resonators \cite{Turner08,OErings,OLrings,Clemmen}, here the integrated intensities, i.e. the total number of pairs detected in transmission, are different for idler and signal. In particular, it is worth noticing that each of the measured intensities is due to the sum of two contributions: photon pairs with both photons exiting on the transmission port and photon pairs in which only one of the photon is transmitted while the other is reflected back (not considering scattering losses). The asymmetry of the structure results in different couplings between the modes and the channels at signal and idler frequencies, thus in Fig. \ref{Fig2}(b) we observe transmission resonances with different heights and quality factors, which can be used for a qualitative interpretation of the results in Fig. \ref{Fig4}. The spontaneous generation rate is  $\sim$300 $\left(\frac{\text{Hz}}{\mu\text{W}^2}\right)$P$_p^2$ with the pump power expressed in $\mu$W, i.e. about two orders of magnitude more than what reported for ring resonators \cite{OLrings,OErings,Clemmen}. Such a large generation rate is essentially due to the smaller volume of PCM as compared to other type of resonators, which indicates this approach as alternative to ring resonators and waveguides.

In conclusion we have reported experimental evidence of stimulated and spontaneous FWM in a silicon-on-insulator photonic molecule. These results indicate that PCMs based on PhC/PhW nano-cavities are very interesting as efficient and compact  sources of classical and non-classical light based on parametric fluorescence. Possible applications range from ultra-low power nonlinear optics to potentially quantum optics integrated on a SOI chip.

This work was supported by CNISM funding through the INNESCO project ``PcPol", by MIUR funding through the FIRB ``Futuro in Ricerca" project RBFR08XMVY, from the foundation Alma Mater Ticinensis and by Fondazione Cariplo through project 2010-0523 "Nanophotonics for thin-film photovoltaics". We aknowledge Lucio Claudio Andreani for fruitful discussion

\end{document}